\documentclass[%
  aps,
  prr,
  preprint,
  amsmath,amssymb,
  superscriptaddress
]{revtex4-2}

\usepackage{graphicx}   
\usepackage{dcolumn}    
\usepackage{bm}         
\usepackage{hyperref}   
\hypersetup{colorlinks=true, linkcolor=blue, citecolor=blue, urlcolor=blue}

\begin{document}
\title{On the White-Noise Limit of the Colored Linear Inverse Model}

\author{Cristian Martinez-Villalobos}
\affiliation{Facultad de Ingeniería y Ciencias, Universidad Adolfo Ibáñez, Santiago, Chile}

\date{\today}

\begin{abstract}
A recent paper by Lien et al.\ (2025)\cite{lien_colored_2025} introduces the ``colored linear inverse model'' (colored LIM), in which stochastic forcing is modeled using Ornstein--Uhlenbeck colored noise rather than idealized white noise. In that work, it is shown that the derivative-based identification formulas used to estimate model parameters do not admit a regular white-noise limit, due to the loss of differentiability of the lag-correlation function at zero lag. Here we revisit the white-noise limit from the perspective of the underlying stochastic differential equations. Treating the colored LIM as an augmented Ornstein--Uhlenbeck system, we show that, as the correlation time $\tau\to 0$, the colored-noise-driven system reduces to the classical LIM, and the corresponding stationary covariance satisfies the standard fluctuation--dissipation relation. Re-examining the same linear system used by Lien et al.\ (2025), we illustrate this convergence numerically. These results highlight a distinction between the singular behavior of derivative-based identification formulas and the regular limiting behavior of the underlying stochastic model. Taken together with recent results showing convergence of estimated parameters in the white-noise limit \cite{lien_JAS_2026}, they provide a consistent interpretation in which the colored LIM recovers the classical LIM at the level of stochastic dynamics, even though certain estimation procedures become ill-defined in that limit.
\end{abstract}

\maketitle

\section{Problem Setup}

A main reason for introducing colored LIMs \cite{lien_colored_2025} is to extend the well-known 
LIM framework \cite{Penland1995_LIM} so that it can represent temporally correlated stochastic forcing. 
In climate dynamics, such forcing often arises from intraseasonal atmospheric variability \cite{gushchina_intraseasonal_2012}, westerly 
wind bursts in ENSO \cite{thual_simple_2016, martinez-villalobos_observed_2019}, or stochastic heat fluxes \cite{sura_daily_2006}. These processes display memory on subseasonal 
to interannual scales. Accounting for this correlation may yield a more realistic view of how unresolved 
variability influences the predictable components of the system.  

For any such extension, it is useful to understand in what sense it connects to the established white-noise theory when the correlation time vanishes. Lien et al.\ (2025) show that the derivative-based correlation formulas used to identify the model parameters do not admit a regular white-noise limit, due to the loss of differentiability of the lag-correlation function at zero lag. In related work, Lien et al.\ \cite{lien_JAS_2026} further show that estimated parameters converge in the white-noise limit. These results reflect a distinction between the behavior of the identification framework and that of the underlying stochastic model. While the derivative-based formulas become singular as $\tau \to 0$, the colored-noise-driven stochastic system itself remains well defined. We therefore revisit the limit directly at the level of the coupled stochastic differential equations and show that the classical LIM is recovered at the level of the stochastic dynamics. In this way, the connection between colored and classical LIM formulations becomes more explicit, while remaining consistent with the behavior of the identification procedures.

It is well established in the stochastic-process literature that Ornstein--Uhlenbeck forcing converges to 
white noise when the correlation time vanishes \cite[e.g.,][]{van_kampen_chapter_2007, gardiner_crispin_stochastic_2009}. 
Our goal is not to re-derive these fundamentals, but to show clearly within the LIM framework how the 
classical formulation is obtained, and to confirm this numerically for the system studied by Lien et al.\ (2025).

The colored LIM considered here is the coupled system
\begin{align}
\frac{d\mathbf{x}}{dt} &= \mathbf{A}\,\mathbf{x} + \mathbf{Q}\,\boldsymbol{\eta}(t), \label{eq:x_eq} \\
\frac{d\boldsymbol{\eta}}{dt} &= -\frac{1}{\tau}\,\boldsymbol{\eta} + \frac{1}{\tau}\,\boldsymbol{\xi}(t), \label{eq:eta_eq}
\end{align}
where $\mathbf{x}(t)\in\mathbb{R}^n$ is the state vector, $\boldsymbol{\eta}(t)\in\mathbb{R}^m$ is a colored-noise process with correlation time $\tau>0$, $\boldsymbol{\xi}(t)$ is standard Gaussian white noise with covariance $\mathbf{I}\,\delta(t-s)$, $\mathbf{A}\in\mathbb{R}^{n\times n}$, and $\mathbf{Q}\in\mathbb{R}^{n\times m}$. In their formulation, Lien et al.\ adopt a single global correlation time 
$\tau$ applied uniformly across components. In principle, however, one could 
allow different variables to have distinct decorrelation times. For realistic 
multivariate systems such as ENSO, this may be more appropriate, since 
sea-surface temperature, thermocline depth, and atmospheric anomalies are 
likely driven by forcings with different memory scales. The behavior discussed by Lien et al.\ in the limit $\tau\to 0$ arises from properties of a derivative-based correlation-function estimator near $s=0$.

\section{Interpreting the Discrepancy}

The classical LIM \cite{Penland1995_LIM} is
\begin{equation}
\frac{d\mathbf{x}}{dt} = \mathbf{A}\,\mathbf{x} + \mathbf{Q}\,\boldsymbol{\xi}(t), \label{eq:white_lim}
\end{equation}
with the fluctuation--dissipation (stationary Lyapunov) relation (e.g., \cite{Penland1994_FDR})
\begin{equation}
\mathbf{A}\,\mathbf{C}_{xx} + \mathbf{C}_{xx}\,\mathbf{A}^\top + \mathbf{Q}\,\mathbf{Q}^\top = \mathbf{0}, \label{eq:FDR_white}
\end{equation}
where $\mathbf{C}_{xx}=\mathbb{E}[\mathbf{x}\mathbf{x}^\top]$ is the stationary covariance. For white-noise forcing, the lag covariance has the closed form
\[
\mathbf{K}(s) \equiv \mathbb{E}[\mathbf{x}(t+s)\mathbf{x}(t)^\top] = e^{\mathbf{A}s}\,\mathbf{C}_{xx}\quad (s\ge 0).
\]

In contrast, the colored-LIM identification procedure of Lien et al.\ is built from the behavior of the correlation function $\mathbf{K}(s)$ near $s=0$, specifically through its first, second, and third derivatives. As they explicitly discuss, this construction becomes singular in the white-noise limit: as $\tau\to 0$, $\mathbf{K}(s)$ loses smoothness at $s=0$, so differentiation and the limiting procedure do not commute. From this perspective, the non-reduction result can be understood as applying to the derivative-based identification formulas, not to the underlying family of colored-noise-driven linear SDEs. In that narrower sense, the two statements are compatible: the estimator does not have a regular white-noise limit, even though the stochastic model itself does.

\section{Recovery of the Classical LIM at the Level of the Stochastic Dynamics}

To avoid conflating the behavior of the identification formulas with that of the stochastic model itself, we now examine the $\tau\to 0$ limit directly for the augmented SDE. Here, recovery at the level of the stochastic dynamics refers to the fact that, as $\tau\rightarrow 0$, the colored-noise-driven system reduces to the classical LIM, implying convergence of statistical quantities such as the stationary covariance.

Consider the augmented linear system for $(\mathbf{x},\boldsymbol{\eta})$:
\begin{equation}
\frac{d}{dt}
\begin{pmatrix}
\mathbf{x} \\
\boldsymbol{\eta}
\end{pmatrix}
=
\underbrace{\begin{pmatrix}
\mathbf{A} & \mathbf{Q} \\
\mathbf{0} & -\tau^{-1}\mathbf{I}
\end{pmatrix}}_{\displaystyle \mathbf{M}}
\begin{pmatrix}
\mathbf{x} \\
\boldsymbol{\eta}
\end{pmatrix}
+
\underbrace{\begin{pmatrix}
\mathbf{0} \\
\tau^{-1}\mathbf{I}
\end{pmatrix}}_{\displaystyle \mathbf{N}}
\boldsymbol{\xi}(t). \label{eq:aug_system}
\end{equation}
This is a multivariate Ornstein--Uhlenbeck process with stationary covariance $\mathbf{C}$ solving
\begin{equation}
\mathbf{M}\,\mathbf{C} + \mathbf{C}\,\mathbf{M}^\top + \mathbf{N}\,\mathbf{N}^\top = \mathbf{0}, \label{eq:aug_lyap}
\end{equation}
where
\[
\mathbf{C}=
\begin{pmatrix}
\mathbf{C}_{xx} & \mathbf{C}_{x\eta} \\
\mathbf{C}_{\eta x} & \mathbf{C}_{\eta\eta}
\end{pmatrix}.
\]

The block equations implied by \eqref{eq:aug_lyap} are:
\begin{align}
\text{(xx)}\quad & \mathbf{A}\,\mathbf{C}_{xx} + \mathbf{C}_{xx}\,\mathbf{A}^\top + \mathbf{Q}\,\mathbf{C}_{\eta x} + \mathbf{C}_{x\eta}\,\mathbf{Q}^\top = \mathbf{0}, \label{eq:xx_block}\\
\text{(x}\eta\text{)}\quad & (\mathbf{A}-\tau^{-1}\mathbf{I})\,\mathbf{C}_{x\eta} + \mathbf{Q}\,\mathbf{C}_{\eta\eta} = \mathbf{0}, \label{eq:xe_block}\\
\text{(}\eta\eta\text{)}\quad & -2\tau^{-1}\,\mathbf{C}_{\eta\eta} + \tau^{-2}\,\mathbf{I} = \mathbf{0}\;\;\Rightarrow\;\;\mathbf{C}_{\eta\eta}=\frac{1}{2\tau}\,\mathbf{I}. \label{eq:ee_block}
\end{align}
From \eqref{eq:xe_block}--\eqref{eq:ee_block} we obtain
\begin{equation}
\mathbf{C}_{x\eta} = (\tau^{-1}\mathbf{I}-\mathbf{A})^{-1}\,\mathbf{Q}\,\mathbf{C}_{\eta\eta}
= \frac{1}{2\tau}\,(\tau^{-1}\mathbf{I}-\mathbf{A})^{-1}\,\mathbf{Q}, \qquad
\mathbf{C}_{\eta x}=\mathbf{C}_{x\eta}^\top. \label{eq:cxeta}
\end{equation}
Substituting \eqref{eq:cxeta} into \eqref{eq:xx_block} gives the exact colored-forcing balance
\begin{equation}
\mathbf{A}\,\mathbf{C}_{xx} + \mathbf{C}_{xx}\,\mathbf{A}^\top
+ \frac{1}{2\tau}\left[
\mathbf{Q}\,(\tau^{-1}\mathbf{I}-\mathbf{A}^\top)^{-1}\,\mathbf{Q}^\top
+ (\tau^{-1}\mathbf{I}-\mathbf{A})^{-1}\,\mathbf{Q}\,\mathbf{Q}^\top
\right]=\mathbf{0}. \label{eq:colored_exact}
\end{equation}
Let $\alpha\equiv\tau^{-1}$. Using the resolvent expansion
$(\alpha\mathbf{I}-\mathbf{A})^{-1}=\alpha^{-1}\mathbf{I}+\mathcal{O}(\alpha^{-2})$ as $\alpha\to\infty$
(i.e., $\tau\to 0$), the bracket in \eqref{eq:colored_exact} equals
\[
\left[\cdots\right]=
\alpha^{-1}\,\mathbf{Q}\,\mathbf{Q}^\top
+\alpha^{-1}\,\mathbf{Q}\,\mathbf{Q}^\top
+\mathcal{O}(\alpha^{-2})
= 2\alpha^{-1}\,\mathbf{Q}\,\mathbf{Q}^\top + \mathcal{O}(\alpha^{-2}).
\]
Multiplying by $\frac{1}{2}\alpha$ yields
\[
\frac{1}{2}\alpha \left[\cdots\right]
= \mathbf{Q}\,\mathbf{Q}^\top + \mathcal{O}(\alpha^{-1}),
\]
so taking $\alpha\to\infty$ in \eqref{eq:colored_exact} recovers the classical LIM relation
\begin{equation}
\mathbf{A}\,\mathbf{C}_{xx} + \mathbf{C}_{xx}\,\mathbf{A}^\top + \mathbf{Q}\,\mathbf{Q}^\top = \mathbf{0}. \label{eq:FDR_limit}
\end{equation}

In the time domain, \eqref{eq:eta_eq} can be written as
\[
\tau\,\frac{d\boldsymbol{\eta}}{dt} = -\boldsymbol{\eta} + \boldsymbol{\xi}(t).
\]
As $\tau\to 0$, the rapidly relaxing OU process satisfies $\boldsymbol{\eta}(t)\Rightarrow \boldsymbol{\xi}(t)$ (convergence in distribution), so \eqref{eq:x_eq} tends to \eqref{eq:white_lim}:
\[
\frac{d\mathbf{x}}{dt} = \mathbf{A}\,\mathbf{x} + \mathbf{Q}\,\boldsymbol{\xi}(t),
\]
i.e., the classical LIM is recovered.

As in \cite{lien_colored_2025}, the derivation above assumed a single global correlation time $\tau$ applied
uniformly to all components of $\boldsymbol{\eta}$. More generally, one may
introduce a diagonal matrix of relaxation times
$\mathbf{T}=\mathrm{diag}(\tau_1,\dots,\tau_m)$ in place of $\tau\mathbf{I}$
in \eqref{eq:eta_eq}. The augmented system
\eqref{eq:aug_system}--\eqref{eq:aug_lyap} then involves
$-\mathbf{T}^{-1}$ on the $\eta$--$\eta$ block of $\mathbf{M}$ and
$\mathbf{T}^{-1}$ in $\mathbf{N}$. The block algebra leading to
\eqref{eq:colored_exact} is unchanged except that the factor
$\tau^{-1}\mathbf{I}$ is replaced by $\mathbf{T}^{-1}$, and the solution of
\eqref{eq:ee_block} becomes
$\mathbf{C}_{\eta\eta}=\tfrac{1}{2}\mathbf{T}^{-1}$. Expanding the resolvent
$(\mathbf{T}^{-1}-\mathbf{A})^{-1}$ for small $\tau_i$ shows that each mode
individually tends to white noise, and the limiting balance again reduces to
the classical LIM relation \eqref{eq:FDR_limit}. Thus the conclusion that the
colored formulation recovers the white LIM as $\tau\to 0$ is robust to
allowing distinct correlation times for different forcing components.

\section{Numerical demonstration using Lien et al.\ (2025) system}

To complement the analytical derivation, we performed a numerical test using
the same three-dimensional linear system presented as equation (32) in
\cite{lien_colored_2025}, up to a normalization convention in the forcing amplitude (a $\sqrt{2}$ factor in their notation).

\begin{equation}
\frac{d\mathbf{x}}{dt}
=
\begin{bmatrix}
-1.0 & 0.5 & 1.0 \\
0.5 & -2.0 & 0.0 \\
0.0 & 1.0 & -1.0
\end{bmatrix}
\mathbf{x}
+
\begin{bmatrix}
1.0 & 0.0 & 0.5 \\
0.0 & 2.0 & 1.0 \\
0.5 & 1.0 & 2.0
\end{bmatrix}
\boldsymbol{\eta},
\label{eq:test_system}
\end{equation}
with $\boldsymbol{\eta}$ governed by \eqref{eq:eta_eq}.

For each choice of correlation time $\tau$ we
constructed the augmented Ornstein--Uhlenbeck system
\eqref{eq:aug_system}--\eqref{eq:aug_lyap} and solved the associated
Lyapunov equation for the stationary covariance. The top-left block
$\mathbf{C}_{xx}(\tau)$ was then compared with the stationary covariance of
the classical white-noise LIM, $\mathbf{C}_{\text{white}}$, obtained from
\eqref{eq:FDR_white}. As a metric we used the Frobenius norm of the
difference, normalized by $\|\mathbf{C}_{\text{white}}\|_F$, namely
\[
\frac{\|\mathbf{C}_{xx}(\tau)-\mathbf{C}_{\text{white}}\|_F}
{\|\mathbf{C}_{\text{white}}\|_F},
\]
where $\|\mathbf{M}\|_F = \sqrt{\sum_{i,j} M_{ij}^2}$ denotes the Frobenius
norm. This norm provides a natural measure of the overall discrepancy between
the two covariance matrices.

\begin{figure}[t]
\centering
\includegraphics[width=0.65\linewidth]{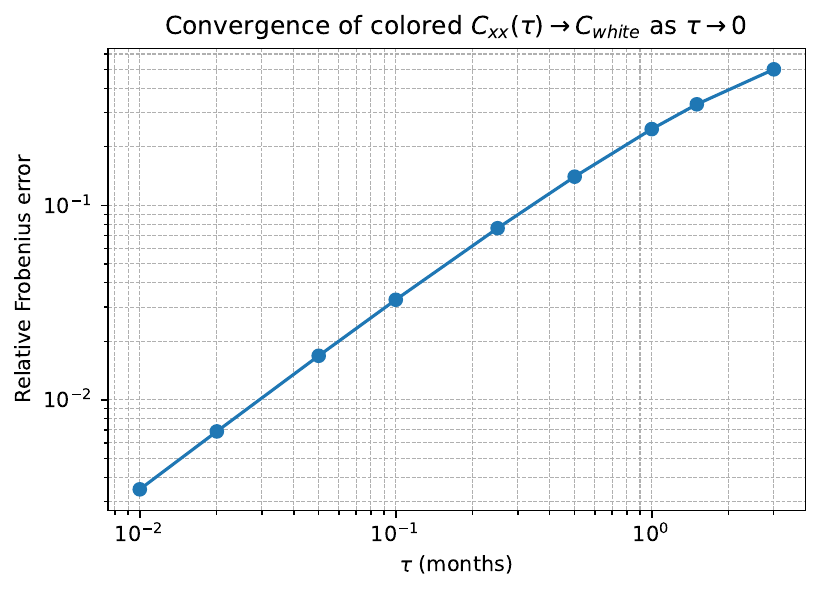}
\caption{Convergence test using the system of \cite{lien_colored_2025},
Eq.~(32). The plot shows the relative Frobenius error
$\|\mathbf{C}_{xx}(\tau)-\mathbf{C}_{\text{white}}\|_F /
 \|\mathbf{C}_{\text{white}}\|_F$
as a function of correlation time $\tau$ (in months) on logarithmic axes.
The error decreases monotonically as $\tau\to 0$, demonstrating recovery of
the classical LIM.}
\label{fig:frob_loglog}
\end{figure}

Figure~\ref{fig:frob_loglog} shows the results of a sweep in $\tau$, starting
from values of order months and decreasing toward zero. As $\tau\to 0$, the
relative Frobenius error decays rapidly, confirming that the stationary
covariance of the colored system converges to that of the classical LIM. The
log--log plot highlights the monotone decrease of the error with $\tau$, with
the discrepancy reaching less than one percent for $\tau=0.01$ months. This
numerical experiment illustrates concretely that, for the underlying augmented stochastic system, the stationary covariance approaches that of the classical white-noise LIM in the small-$\tau$ limit. It does not imply that the derivative-based identification formulas themselves remain regular in that limit.

\section{Conclusions}

The analysis presented here shows that the colored-noise-driven linear system underlying the colored LIM recovers the classical LIM in the white-noise limit when that limit is taken directly at the level of the stochastic dynamics. By contrast, the derivative-based identification formulas emphasized by Lien et al.\ become singular as $\tau \to 0$, because the lag-correlation function loses differentiability at zero lag. These results highlight a distinction between the behavior of the identification framework and that of the underlying stochastic model. Taken together with recent results showing convergence of estimated parameters in the white-noise limit \cite{lien_JAS_2026}, they support a consistent interpretation in which the colored LIM and the classical LIM are connected through a regular limit at the level of stochastic dynamics, even though certain estimation procedures do not themselves admit a regular white-noise limit.

\begin{acknowledgments}
We acknowledge support from Proyecto ANID Fondecyt Iniciación 11250471. The author thanks Justin Lien for helpful and constructive comments on an earlier version of this note.
\end{acknowledgments}

\bibliography{Colored_LIM}

\end{document}